\magnification \magstep2
\def \etal {{\it et al.}}
\newcount \secnum
\newcount \eqqqnum
\def \section #1 {\vskip \baselineskip \goodbreak \advance \secnum by 1 \leftline{\bf \the\secnum \ #1 \hfil}  \par }
\font \author=cmssi12
\font \institute=cmssi10
\font \title=cmss17
\nopagenumbers

\centerline{\title Combining Experiments with Systematic Errors}

\vskip 1 cm

\centerline{\author Roger Barlow}

\vskip .5 cm
\centerline{\institute The University of Huddersfield, Huddersfield HD1 3DH, UK}
\vskip 1 cm

\centerline{\institute $26^{th}$ September 2020}
\vskip 2 cm
\centerline{\it Abstract}
\vskip 1 cm

\centerline{\vbox{\hsize 8 cm 
\noindent We consider fits to two or more datasets for which results from the same experiment share a common systematic uncertainty in addition to their individual statistical errors. 
This is important in extracting the maximum information from a set of similar but different  experiments (or the same experiment under different conditions)  analysing similar but different datasets, as happens at the LHC and other particle colliders. 
There are two techniques in use: using the full matrix and using
extra paramneters,  and we show, for a completely general fit, that for an additive uncertainty they are in principle equivalent even though in practice the details differ. For a multiplicative error 
the matrix fit is equivalent to the extra parameter fit if the factor is applied to the data points but not if it is applied to the function: the former
leads to biassed estimates and the latter avoids them.  }}

\vfill
\eject

\section {The problem}

\footline={\hss -- \folio -- \hss}

It is frequently required to fit the parameter(s) $a$ of a  function $f(x;a)$ 
to values $\{x_i,y_i\}$ where the measurements come from several different 
experiments;  a measurement $i$ in experiment $E_r$ is subject not only to an independent error $\sigma_i$ but to a systematic uncertainty $S_r$ which is
shared with all other measurements of that experiment.  
This uncertainty may be additive (for example, in a common background), or multiplicative (for example, in a common efficiency).

For example the $e^+e^-$ hadronic  cross section ratio  $R$ has been measured by many experiments over a wide range of energies and
fitting parameters, for example the $Z$ width [1], is best done by a global fit to the different experiments.
Many determinations of neutrino mixing parameters, for example [2], involve global fits to 
several different neutrino experiments.
Describing proton-proton elastic and diffractive scattering using Pomeron exchange models 
[3] fits parameters using data from many diffierent experiments at the 
ISR, the Tevatron and the LHC with each experiment sharing a systematic uncertainty.  

\vskip 5.5 cm
\includegraphics{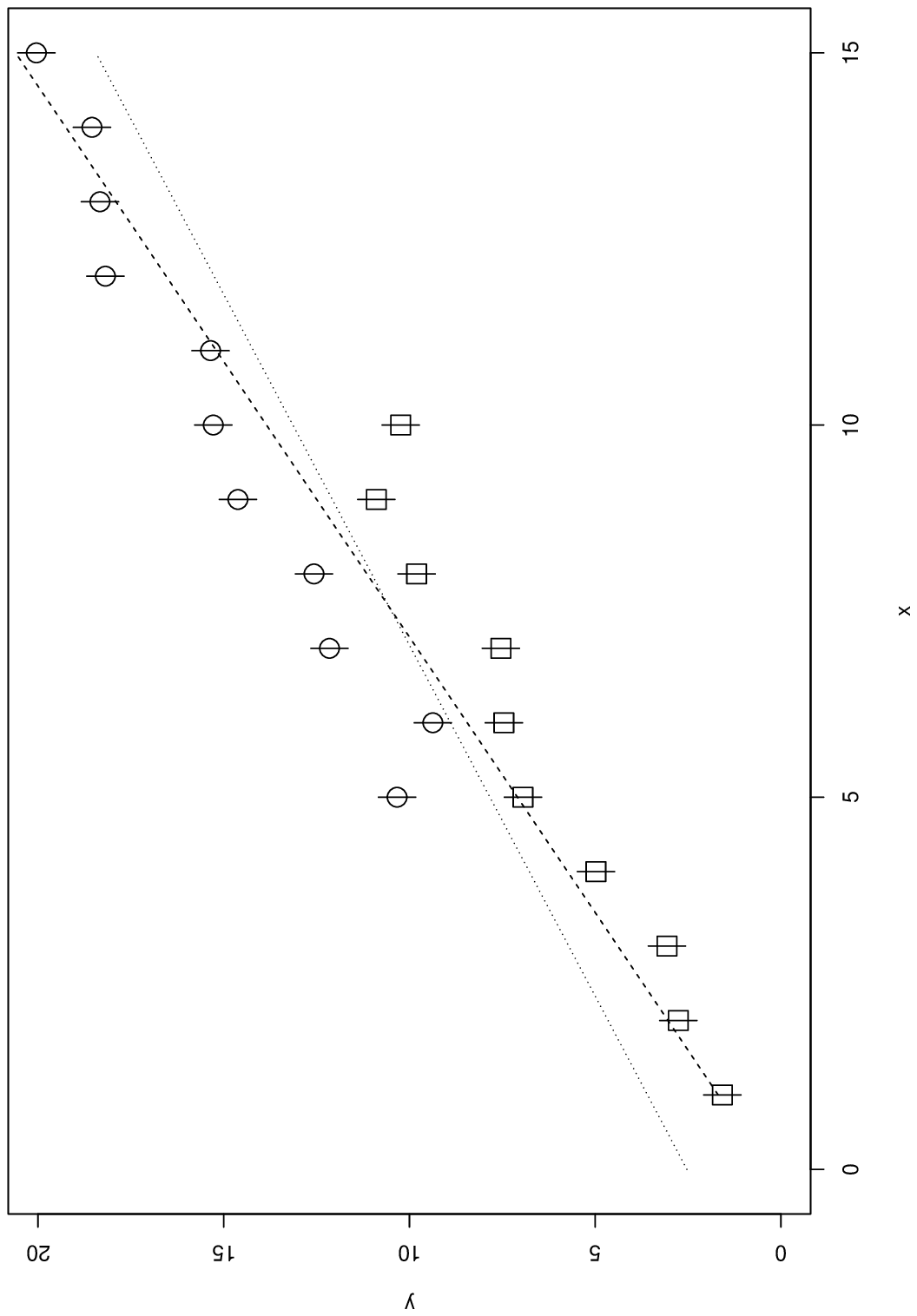}

\centerline{Figure 1: Data from two different experiments}

An illustration is shown in Figure 1, where straight-line-fit data from two different experiments is 
depicted by circles and squares. There is a clear systematic shift between the two datasets. Each 
on its own is well described by a straight line, but if the two are combined without regard for the systematic the dashed line is obtained, which is clearly wrong. The dotted line is produced by the techniques discussed below.

The problem is not that there is no obvious way to perform such a fit: the 
problem is that there are two such methods: by full matrix inversion and by 
introducing additional parameters.

Suppose that there are $n$ experiments, $E_1 \dots E_r \dots E_n$.  (In what follows we use the index $r$ to denote the experiment, and indices $i,j,k$ to denote individual data points.)
For uncorrelated data points one determines the value(s) of $a$ by minimising the $\chi^2$
$$ \chi^2=\sum_i \left({y_i-f(x_i;a)\over \sigma_i}\right)^2 \eqno(1) $$ 
For correlated data the $1/\sigma^2$ terms are replaced by
the inverse of the error matrix

$$ \chi^2=\sum_{i,j} \left(y_i-f(x_i;a)\right)V^{-1}_{ij}\left(y_j-f(x_j;a)\right)  \eqno(2)$$
and the value(s) of $a$ are determined by minimising this.

However alternatively one can introduce parameters $z_r$ to describe the shared deviations of each experiment. For an additive uncertainty
 $z_r$ has mean zero and standard deviation $S_r$, and the $\chi^2$ is written as

$$
\chi^2=\sum_r \sum_{i \in E_r} \left( {y_i-f(x_i;a)-z_r \over \sigma_i}\right)^2 + \sum_r \left( {z_r \over S_r}\right)^2
\eqno(3)
$$

If the uncertainty is multiplicative then 
the $\chi^2$ can be written

$$
\chi^2=\sum_r \sum_{i \in E_r} \left( {y_i-z_r f(x_i;a) \over \sigma_i}\right)^2 + \sum_r \left( {z_r -1  \over S_r}\right)^2
\eqno(4)
$$
where the $z_r$ scale factors have mean 1.0.

We shall show that these apparently different fitting procedures lead to the same result.

\section {Past Approaches}

D'Agostini [4] discusses the structure of the covariance matrix for muliplicative 
systematic uncertainties. In considering the special case of forming the average of two quantities (which 
corresponds to $f(x;a)=a)$) they assert that the  methods give the same result,
but do not generalise this statement.  
Demortier, in an unpublished note [5], shows the methods are equivalent for 
additive uncertainties, giving quite a complicated proof. Mo [6] considers multplicative uncertainties 
but only considers linear fits, for which they assert that the methods are equivalent (they refer to a paper which is unfortunately only in Chinese.) 
Blobel [7] 
asserts that the two approaches are equivalent ``for simple problems'' and
``for additive uncertainties''  but does not offer a proof.
Fogli \etal [8] give a proof of equivalence for the additive case. 

\section {Additive Systematic Errors}

The inversion of the matrix $V$ can be performed algebraically.
First, note that  if the data are grouped together according to their experiment, $V$ is block-diagonal, each block having the  form 
 
$$
V_r=
\left(
\matrix{
\sigma_1^2+S_r^2 & S_r^2 & S_r^2 & \dots \cr
S_r^2 & \sigma_2^2 + S_r^2 & S_r^2 & \dots \cr
S_r^2 & S_r^2 & \sigma_3^2+S_r^2 & \dots \cr
\vdots &\vdots &\vdots  \cr
}
\right)
\eqno(5)
$$
so the inverse of $V$ is also block diagonal in the individual inverse 
matrices $V_r^{-1}$, and Equation~(2) has the form
$$
\chi^2 = \sum_r \sum_{i\in E_r} \sum_{j\in E_r} \Delta_i {(V_r^{-1})}_{ij} \Delta_j
\eqno(6)
$$
where the sums over $i$ and $j$ run over measurements of the experiment $r$ and
$\Delta_i = y_i-f(x_i;a)$.

The individual submatrices can also be inverted. 
To see this, we first consider the case where all $\sigma_i$ are the same. 
Introduce the matrix $U$ for which all members are unity, and note that for an $n \times n$ square matrix, 
$U^2=n U$.
The submatrix may be written as $V_r=\sigma^2 I + S_r^2 U$, and $V^{-1}_r$
is just: 
$$
V_r ^{-1}= {1 \over \sigma^2} I - \left({S_r^2 \over \sigma^2 (\sigma^2+n_r S_r^2)}\right) U 
\eqno(7)
$$

If the $\sigma_i$ are different then the situation is not quite so elegant, but the inverse is
 still writeable in closed form as 
$$
{(V_r ^{-1})}_{ij}= {\delta_{ij} \over \sigma_i^2} 
 - {S_r^2 \over 1 + \sum_{k\in E_r} {S_r^2 \over\sigma_k^2}}{1 \over \sigma_i^2 \sigma_j^2} 
\eqno(8)
$$
as can be seen  by multiplying out this expression and $(V_r)_{ij}=\delta_{ij}\sigma_i^2+S_r^2$.

The $\chi^2$ to be minimised in the first method is thus

$$
\chi^2=\sum_r \sum_{i\in E_r}  {\Delta_i^2 \over \sigma_i^2} - \sum_r {S_r^2 \over 1
+ \sum_{k \in E_r} {S_r^2 \over\sigma_k^2}} 
\sum_{i\in E_r} \sum_{j\in E_r} {\Delta_i \Delta_j \over \sigma_i^2 \sigma_j^2} .
\eqno(9)
$$

For the second method the expression for $\chi^2$ is given by Equation~3.
To find the minimum the differentials are set to to zero, and the resulting
system of simultaneous equations partially decouples.    The differential for one of the $z_r$ is

$$
{\partial \chi^2 \over \partial z_r}= -2\sum_j {\Delta_j-z_r \over \sigma_j^2}+2 {z_r\over S_r^2}
\eqno(10)
$$
which gives the estimate of $z_r$  for a particular $a$, writing, for convenience, $\zeta_r= {1 \over {1 \over S_r ^2}+\sum_k {1 \over \sigma_k^2}}$

$$
\hat z_r = \zeta_r \sum_j {\Delta_j \over \sigma_j^2}.
\eqno(11)
$$
Inserting Equation 11 into  Equation 3  gives

$$\eqalign{
	\chi^2=\sum_r \Biggl( \Biggr. &
 \sum_{i\in E_r} {\Delta_i^2 \over \sigma_i^2}-2 \sum_{i\in E_r} {\Delta_i \over \sigma_i^2 } \zeta_r \sum_{j\in E_r} {\Delta_j \over \sigma_j^2}
 \cr &
 +\left( \sum_{k\in E_r}{1 \over \sigma_k^2}  + {1 \over S^2}\right) \zeta_r^2 \sum_{i\in E_r} \sum_{j\in E_r} {\Delta_i \over \sigma_i^2}{\Delta_j \over \sigma_j^2}
\Biggl. \Biggr)
 }
 \eqno(12)
$$
which reduces to the expression in Equation 9.

 The two methods are thus equivalent, minimising the same $\chi^2$. 
 (This is used to obtain the dotted line in Figure~1.)
From the mathematical point of view this 
may be surprising as the formul\ae\  
are very different; it is gratifying from the statistical viewpoint as 
both approaches make the same assumptions.
  
  Either method can therefore be chosen in a particular problem.
The matrix method is simpler, particularly if the model $f(x;a)$ is linear in $a$ as the normal equations can be solved directly, without the need for iteration.
However the second method gives the $z_r$ offsets as a by-product, and these may be
useful to investigate the quality of the different datasets.

The second method increases the difficulty of the minimisation by
  more than just having extra parameters to fit. If  the function includes an additive constant term
  (which it generally will),  the search space 
  includes a
 direction in which a change in this term can be balanced by an opposite shift in the mean $z$. In this direction  $\chi^2$  changes slowly, through the $1/S^2$ terms,
 whereas the change in any orthogonal direction is, assuming that there are many more
 datapoints than datasets, much larger. This causes a problem of the Rosenbrock's Valley type.
 Even if the minimiser reports `successful convergence', the result should be 
confirmed  with care. It may be 
preferable, rather than setting $a$ and all the $z_r$ as adjustable parameters in
the $\chi^2$ of Equation~3, to work
solely using the $a$ parameters, with the $z_r$ at every step being given explicitly by Equation~11.
 
\section {Multiplicative Systematic Errors}

Correlated multiplicative errors are more common than correlated additive errors, as
uncertainties in acceptance, efficiency and overall normalisation are of this type. 
 
 Again one can use a matrix method or introduce additional parameters. The submatrix for individual experiment $E_r$ which has a fractional uncertainty $\xi_r$  is $(V_r)_{ij} = \delta_{ij} \sigma_i^2 + \xi^2_r   y_i y_j$,
for which the block diagonal  inverse, equivalent to Equation~8, is
$$
(V_r)^{-1}_{ij} = {\delta_{ij} \over \sigma_i^2}-{\xi_r^2 y_i y_j \over 1+ \xi_r^2 \sum_{k\in E_r} {y_k^2 \over \sigma_k^2}}\left({1 \over \sigma_i^2 \sigma_j^2} \right) 
\eqno(13)
$$.

For the separate fit, there is an ambiguity about whether to apply the error to the value or the function.
In comparing some measured $y_i$ values with function values $f(x_i,a)$ with a nominal fractional error $\xi$ one can  apply scale factors $z_r$ and write  the $\chi^2$ as

$$
\chi^2=\sum_r \sum_{i\in E_r} \left( {z_r y_i -   f_i \over \sigma_i} \right)^2 + \sum_r \left( {z_r-1 \over \xi_r}\right)^2
\eqno(14)
$$
or as
$$
\chi^2=\sum_r \sum_{i\in E_r} \left( {y_i - z_r f_i \over \sigma_i} \right)^2 + \sum_r \left( {z_r-1 \over \xi_r}\right)^2
\eqno(15)
$$
	
	Leaving aside for now the arguments for the choice  -  which amounts to the difference between whether a quantity or its reciprocal is better described by a Gaussian, a question that does not arise for an additive correction - one can estimate the $z_r$ for a given $a$ in the first case by 
setting the differential of  
 Equation~14  to zero
$$
 \hat z_r={1 + \xi_r^2 \sum_{i\in E_r}  {f_i y_i \over \sigma_i^2}  \over 1+  \xi_r^2\sum_{i\in E_r}{y_i^2 \over \sigma_i^2} }
\eqno(16)
$$
and if these are inserted into Equation 14 then this gives 

$$\chi^2=\sum_i {\Delta_i^2 \over \sigma_i^2} -
\sum_r {\xi_r^2 \over 1+\xi_r^2 \sum_{k \in E_r} {y_k^2 \over \sigma_k^2}} \left( \sum_{i \in E_r}{\Delta_i y_i \over \sigma_i^2}\right)^2\eqno(17)$$  
which exactly matches the $\chi^2$ given by the matrix elements in Equation~13.  So the methods are equivalent.

For the second case one differentiates Equation~15
to obtain
$$
 \hat z_r={1 + \xi_r^2 \sum_{i\in E_r}  {f_i y_i \over \sigma_i^2}  \over 1+  \xi_r^2\sum_{i\in E_r}{f_i^2 \over \sigma_i^2} }
\eqno(18)
$$
Inserting this into Equation~15 gives
$$
\chi^2= \sum_i {\Delta^2_{i} \over \sigma_i^2}
-\sum_r   {\xi_r^2 \over 1+ \xi_r^2 \sum_{k \in E_r}  {f_k^2 \over \sigma_k^2}}\left(\sum_{i \in E_r} {\Delta_i f_i \over \sigma_i^2} \right)^2
\eqno(19)
$$
which is similar to Equation~13 except that the measured $y_i$ have been replaced by
the predicted $f_i$ everywhere except in the differences $\Delta_i  = y_i - f_i$

Now consider which of Equations~14 and 15 should be used.
 In general the second choice, applying the error to the function, will be correct. If one measures a value $x=100.0$ with 10\% precision one writes it
as $100.0 \pm 10.0$ for convenience, but this conceals the small difference between a true value of
110.0, for which the deviation is slightly under one sigma, and a true value of 90.0 for which it is slightly over.   
The situation is analogous to the choice between minimising 
$\sum{(n_i-f_i)^2 / f_i}$
or 
$\sum{(n_i-f_i)^2 / n_i}$
in a least squares fit to a histogram: the latter is quicker but leads to a bias which is noticeable for small $n_i$.

\section {Remarks on bias}

That the matrix method, applying the 
multiplicative errors to the values,
leads to a bias was pointed out by D'Agostini[4].

He gives a simple example in which the effect can be seen. Consider just two values with a 
shared multiplicative error $\xi$ and one just seeks
a combined best value. 
A covariance matrix is constructed according to $V_{ij}=\delta_{ij} \sigma^2+\xi^2 x_1 x_2$
and inverted to form the $\chi^2$ to be minimised. 

 $$
\chi^2= (x_1-\hat x,x_2-\hat x)\left( \matrix{x_1^2 \xi^2+\sigma_1^2&x_1 x_2 \xi^2\cr   x_1 x_2  \xi^2& x_2^2  \xi^2+\sigma_2^2)\cr} \right)^{-1} \left( \matrix{x_1-\hat x\cr x_2-\hat x}\right)
	 \eqno(20)
	 $$
 The 
$2 \times 2$ matrix can be inverted explicitly. Finding the maximum 
by differentiating and setting to zero gives
$$
\hat x = { x_1 \sigma_2^2 + x_2 \sigma_1^2 \over \sigma_1^2 + \sigma_2^2 
+ (x_1-x_2)^2 \xi^2}
	 \eqno(21)
$$

The third term in the denominator clearly leads to bias in the estimator
as it is positive definite and makes $\hat x$ smaller than the unbiassed weighted sum. 
This is illustrated in Figure~2 which shows results from 100,000 simulations of averaging two
numbers, both with mean value 1 and individual errors 0.1, with various values 
for the multiplicative error $\xi$.  As the multiplicative error increases 
the spread in the results increases, as one would expect, but also a definite 
bias emerges. the mean is less than 1 and the distribution of results has a negative skew.
\vfill\eject

\
 
\vskip 4.9 cm
\includegraphics{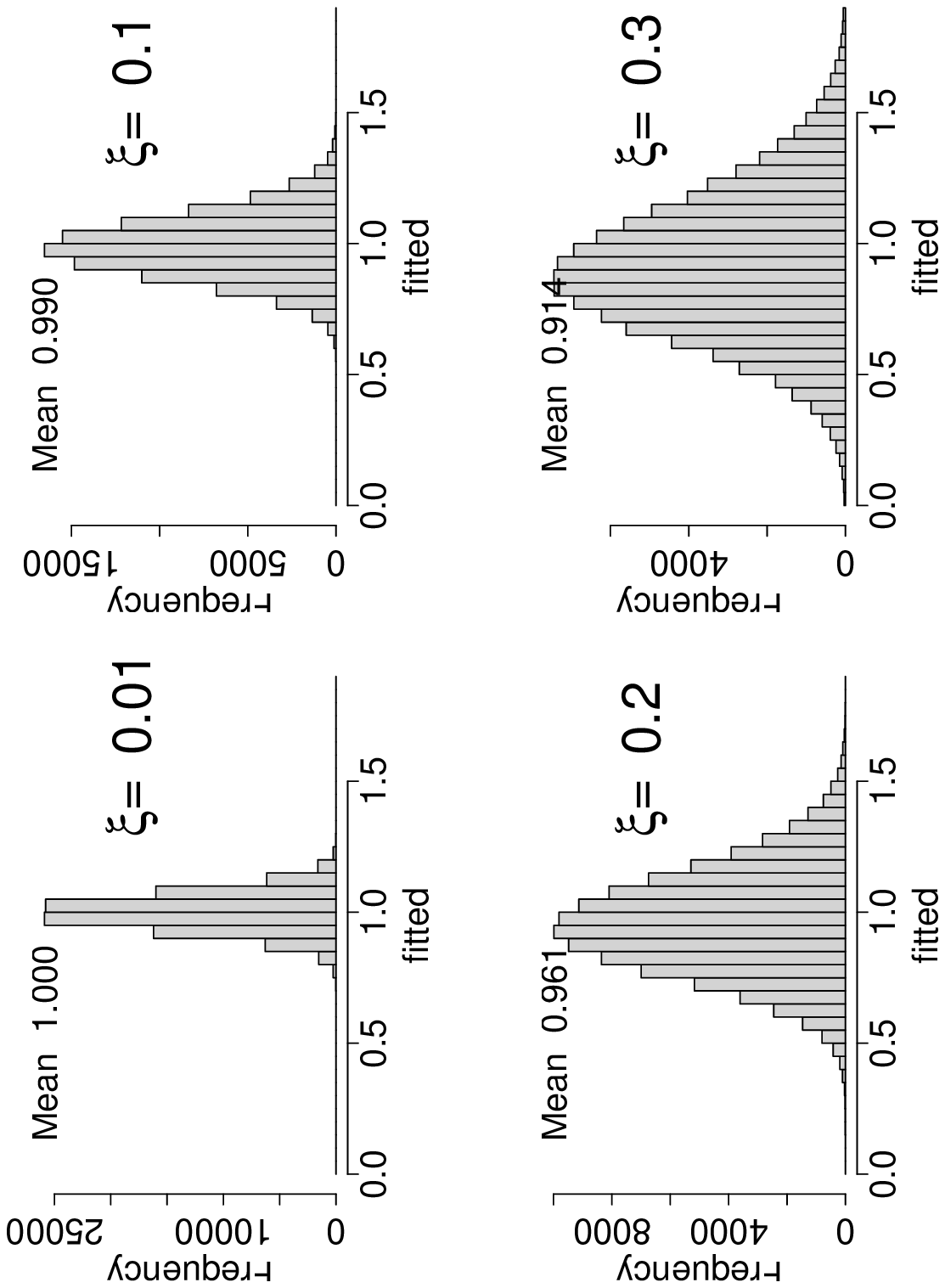}

\includegraphics{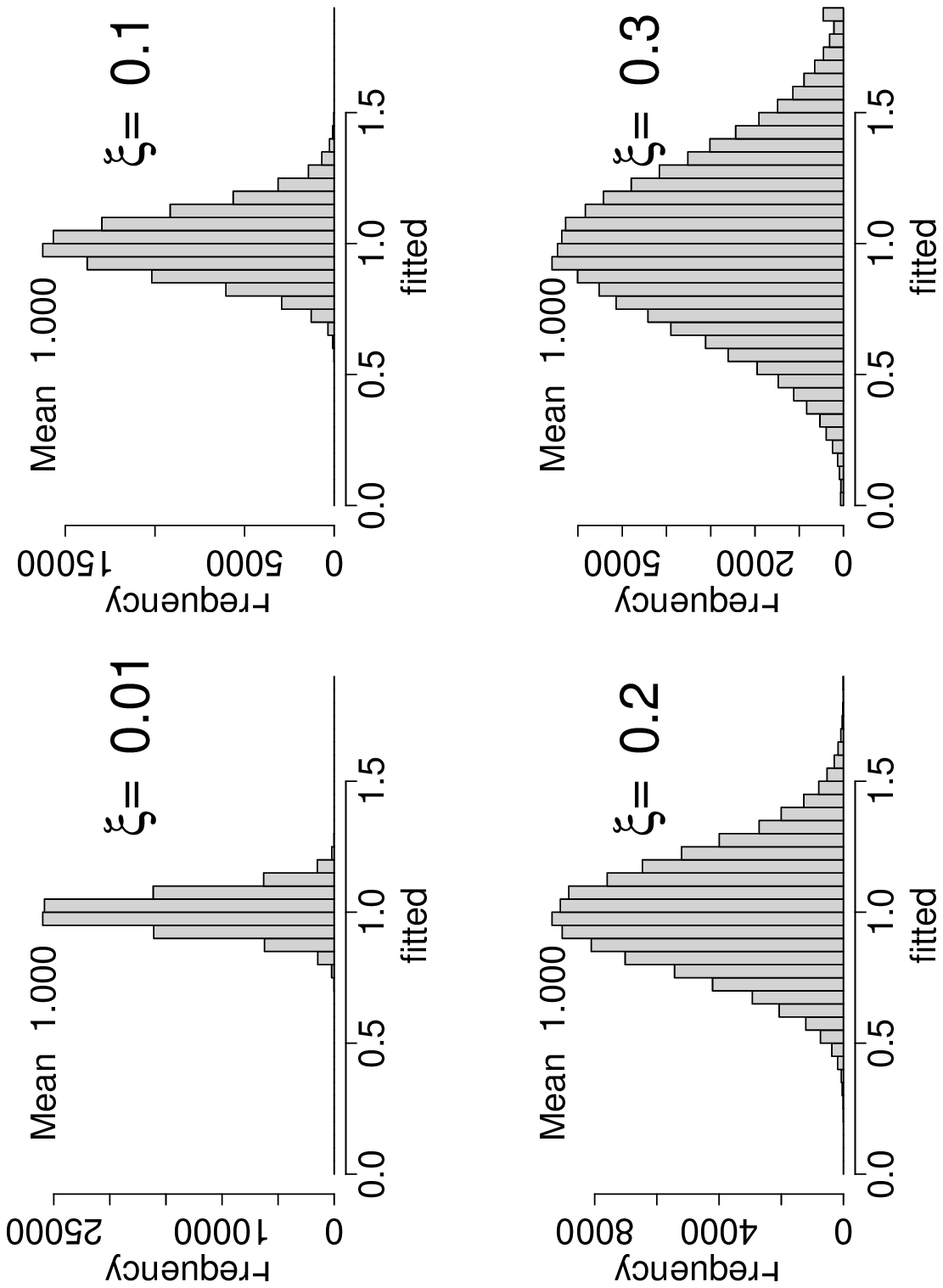}

\centerline{\vbox{Figure 2:
Bias for the average of two numbers with different values of the shared 
systematic $\xi$. The results on the left are using  Equation~21 and on the right using  Equation~22. }
}

\

This bias arises from the scaling, rather than the correlation.
Indeed, if 
 it were not for the correlation then the term that produces the bias would be larger:  $(x_1^2+x_2^2) \xi^2$.  It is a familiar effect when
downward and upward fluctuations are (wrongly) assigned errors that are smaller and larger, respectively. The correlation changes this to   $(x_1-x_2)^2 \xi^2$:
if the measurements happen to be the same the effect vanishes.

The alternative is to apply the factor to the result and minimise

$$
\chi^2= (x_1-\hat x,x_2-\hat x)\left( \matrix{\hat x^2 \xi^2+\sigma_1^2&\hat x^2\xi^2\cr \hat x^2 \xi^2& \hat x^2 \xi^2+\sigma_2^2)\cr} \right)^{-1} \left( \matrix{x_1-\hat x\cr x_2-\hat x}\right)
\eqno(22)
$$
which gives an unbiassed result, as also shown in Figure~2.  This takes noticeably longer to solve as $\hat x$ has to be found by iteration rather than a neat closed form solution, but this is 
usually a price worth paying.

\section Conclusions

The matrix inversion method and the additional-parameter method have been 
shown to be equivalent, for both additive and multiplicative systematic errors. 
Although this has been asserted and proved in some cases in the past, we have shown that it holds in general. 
This result deserves to be widely known, as means that those performing such fits 
can adopt whichever method they prefer 
without deliberating over which is more `correct' in their case.

With either method, introducing multiplicative errors will lead to a bias in the
result if they are applied to the measurements, but this can readily be avoided by
applying the factors to the fitted values. This will make a linear problem become non-linear, but this is unlikely to present a real drawback.

\

\leftline{\bf References}
\parindent 0 pt
\parskip 3 pt

[1] V. Branchina \etal, {\it Hadronic Width of the $Z$ from a Global Fit to $e^+e^-\to$ hadrons data}, Phys. Rev. Lett. {\bf 65} 3237 (1990)

[2] G.L. Fogli \etal, {\it Hints of $\theta_{13}>0$ from Global Neutrino Data Analysis},
Phys. Rev. Lett {\bf 101} 141801 (2008)

[3] R. B. Appleby \etal, {\it The Practical Pomeron for High Energy Proton Collimation}, Eur Phys. J. {\bf C76} 520 (2016)

[4] G. D'Agostini, {\it On the use of the covariance matrix to fit correlated data},
Nucl. Instr. \& Meth. A {\bf 346} 306 (1994)

[5] L Demortier {\it Equivalence of the best-fit and covariance-matrix methods for comparing binned data with a model in the presence of correlated systematic uncertainties}, note  CDF/MEMO/STATISTICS/PUBLIC/8661 (1999)

[6] X-H Mo, {\it Unbiased $\chi^2$ Estimator for linear function fit involving correlated data}, HEP \& NP {\bf 31} 745 (2007)

[7] V Blobel, {\it Data Combination in Particle Physics},
Terascale School on data combination and Limit Setting, DESY, unpublished (2011)

[8] G L Fogli \etal,{\it Getting the most from the statistical analysis of solar neutrino oscillations}
Phys. Rev. {\bf D66} 05310 (2002)
\bye